\documentclass[twocolumn]{aastex61}
\pdfoutput=1 
\usepackage{amsmath,amstext}
\usepackage[T1]{fontenc}
\usepackage{apjfonts} 
\usepackage[figure,figure*]{hypcap}

\newcommand{\kms}{\ensuremath{{\rm km}\,{\rm s}^{-1}}}

\newcommand{\msol}{M$_\odot$}

\shorttitle{Binary dwarf carbon star}
\shortauthors{Margon et al.}

\begin{document}

\title{THE BINARY DWARF CARBON STAR SDSS J125017.90+252427.6}

\author{Bruce Margon}
\affiliation{Department of Astronomy \& Astrophysics, University of California, Santa Cruz, CA 95064, USA}
\author{Thomas Kupfer}
\affiliation{Division of Physics, Mathematics and Astronomy, California Institute of Technology, Pasadena, CA 91125, USA}
\author{Kevin Burdge}
\affiliation{Division of Physics, Mathematics and Astronomy, California Institute of Technology, Pasadena, CA 91125, USA}
\author{Thomas A. Prince}
\affiliation{Division of Physics, Mathematics and Astronomy, California Institute of Technology, Pasadena, CA 91125, USA}
\author{Shrinivas R. Kulkarni}
\affiliation{Division of Physics, Mathematics and Astronomy, California Institute of Technology, Pasadena, CA 91125, USA}
\author{David L. Shupe}
\affiliation{Infrared Processing and Analysis Center, California Institute of Technology, Pasadena, CA 91125, USA}

\begin{abstract}
Although dwarf carbon (dC) stars are thought universally to be binaries to explain the presence of $C_2$ in their spectra while still near main sequence luminosity, direct observational evidence for binarity is remarkably scarce. Here we report the detection of a 2.92 d periodicity in both photometry and radial velocity of SDSS J125017.90+252427.6, an $r=16.4$ dC star. This is the first photometric binary dC, and only the second dC spectroscopic binary. The relative phase of the photometric period  to the spectroscopic observations suggests that the photometric variations are a reflection effect due to heating from an unseen companion. The observed radial velocity amplitude of the dC component ($K = 98.8\pm10.7$~\kms) is consistent with a white dwarf companion, presumably the evolved star that earlier donated the carbon to the dC, although substantial orbital evolution must have occurred. Large synoptic photometric surveys such as the Palomar Transient Factory, used for this work, may prove useful for identifying binaries among the shorter period dC stars.
\end{abstract}

\keywords{stars: carbon --- binaries: spectroscopic}

\section{Introduction}
Although the first dwarf carbon (dC) star was identified 40 years ago \citep{1977ApJ...216..757D}, these objects, characterized by prominent $C_2$ absorption bands, yet main sequence luminosity, remain enigmatic in many ways. The excellent review by \citet{2013ApJ...765...12G} (hereafter "G13") points out that the only plausible source of photospheric $C_2$ prior to post main sequence evolution is mass transfer from a faster evolving but now inconspicuous binary companion.  Aside from the origin of the $C_2$ features, dC stars also pose other interesting problems. The prototype, G77-61, is extraordinary metal poor \citep{2005A&A...434.1117P}. Recently several very high velocity dC stars have also been suggested as candidates for ejecta from binary supernova \citep{2016ApJ...833..232P}.

Despite wide agreement that dC stars must be (or have been) binaries, direct evidence for the binary nature of these objects is remarkably scarce.  Thanks chiefly to the Sloan Digital Sky Survey (SDSS) \citep{2000AJ....120.1579Y}, there are now close to $10^3$ dC stars identified \citep{2002AJ....124.1651M, 2004AJ....127.2838D, 2013ApJ...765...12G, 2014RAA....15.1671S}, yet only barely more than a handful are directly observed to be binaries. Remarkably, there is only one spectroscopic binary, the prototype dC star G77-61 \citep{1986ApJ...300..314D}. About a dozen are known to have dC/DA white dwarf composite spectra, and $\sim20$ show emission lines (G13 and references therein), suggestive of membership in a binary, but perhaps instead due to chromospheric activity on a single star. Only the (one) spectroscopic binary is amenable to even begin quantitative characterization of the system. In this respect dC stars frustratingly differ from a variety of other post-mass transfer systems, e.g., Barium stars, where essentially all are observed to be binary \citep{1980ApJ...238L..35M}. There are very few papers in the literature reporting negative results in searches for dC radial velocity variations (cf. \citet{1994ApJ...423..723G}), so it is difficult to judge how extensively this problem has been pursued spectroscopically.

Here we discuss a new approach to identifying binaries among the dC stars, and report on the first successful example.

\section{Observations}
\subsection{Photometry}
We utilized the Palomar Transient Factory (PTF) \citep{2009PASP..121.1395L, 2009PASP..121.1334R} to search for photometric periodicities in the list of SDSS faint high latitude carbon stars presented by G13 and \citet{2014RAA....15.1671S}. PTF data reduction is described in \citet{2014PASP..126..674L},
while the photometric system is discussed in \citet{2012PASP..124...62O}.
The relative photometric calibration uses the prescription of \citet{2011ApJ...740...65O}. Approximately $10^{3}$ objects were examined for periods in the range $10^{-1}$ - $10^{3}$ day. We used the Vartools Lomb-Scargle algorithm \citep{2016A&C....17....1H} to identify the most probable periods in the lightcurve.

Although several objects with promising periodicities appeared in this analysis, one object in particular displayed periodic behavior with very high probability, and thus was chosen for further study. The $r=16.4$ dC star SDSS J125017.90+252427.6 (hereafter "J1250") displays a period $P=2.92217\pm0.0015$\,days, with $r$-band semi-amplitude $0.023\pm0.002$\,mag, and a formal false occurrence probability, calculated by the \citet{2016A&C....17....1H} formalism, of $\sim10^{-21}$. The uncertainty on the derived period was estimated as described by \citet{2013MNRAS.432.2048K},
using a simple Monte Carlo simulation, where $10^3$ periodograms were computed, with each iteration deleting a random number of randomly chosen data points, which are then replaced with a duplicate of a different observation. For each such simulated data set, the highest periodogram peak was taken as the orbital period, and the standard deviation of the ensemble of these trial peaks is adopted as the final uncertainty. 

In Figure~1 we show a periodogram of analysis of the $r$-band PTF data on J1250, consisting of 173 observations obtained with random cadence over an interval of almost 6 years. Figure~2 displays a folded light curve at the best fit sine wave period, together with residuals from this fit. The phase of the sine wave was fixed, and was obtained from the best-fit radial velocity curve ($\S$2.2). Examination of the window function (i.e., an identical analysis of an artifical pure sine wave evaluated at the same times as the actual data) reveals that the observed higher harmonics in Figure 1 are not necessarily informative on the shape of the light curve, but rather result from a finite length data set of poorly spaced observations of an underlying period close to an even mutiple of one day.

\begin{figure}
\begin{center}
\includegraphics[width=0.49\textwidth]{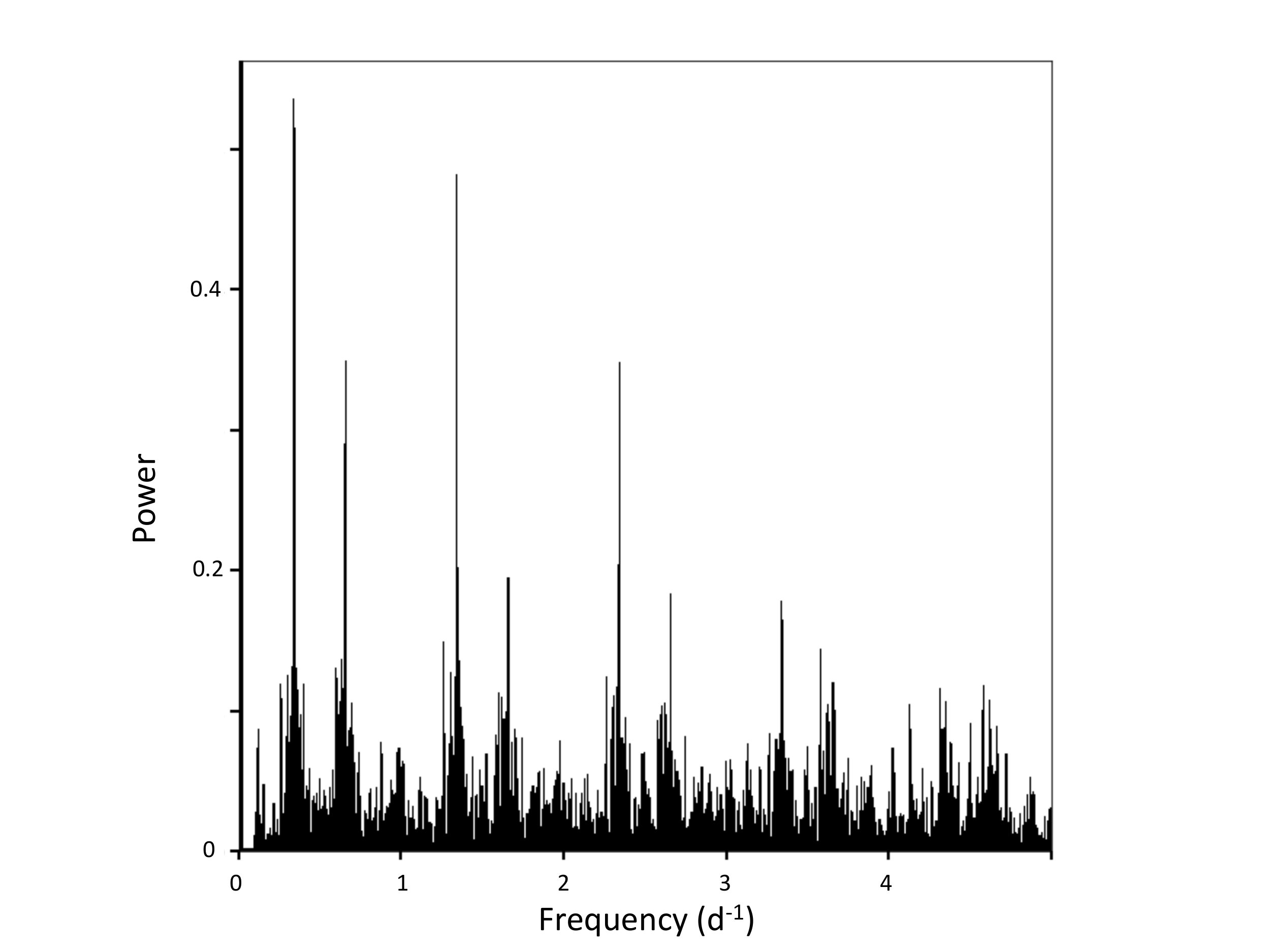}
\caption{Periodogram of  PTF $r$-band photometry of J1250, derived via the Lomb-Scargle algorithm.
The power is an empirical overlap of the periodic signal with a sinusoid at the same frequency, where unity represents perfect overlap. The strong detection of a period at 0.342~d$^{-1}$ and higher harmonics is readily apparent. A second set of peaks at 0.056~d$^{-1}$ is the beat of this primary period with 1 day.}
\label{fig:periodogram}
\end{center}
\end{figure}

\begin{figure}
\begin{center}
\includegraphics[width=0.49\textwidth]{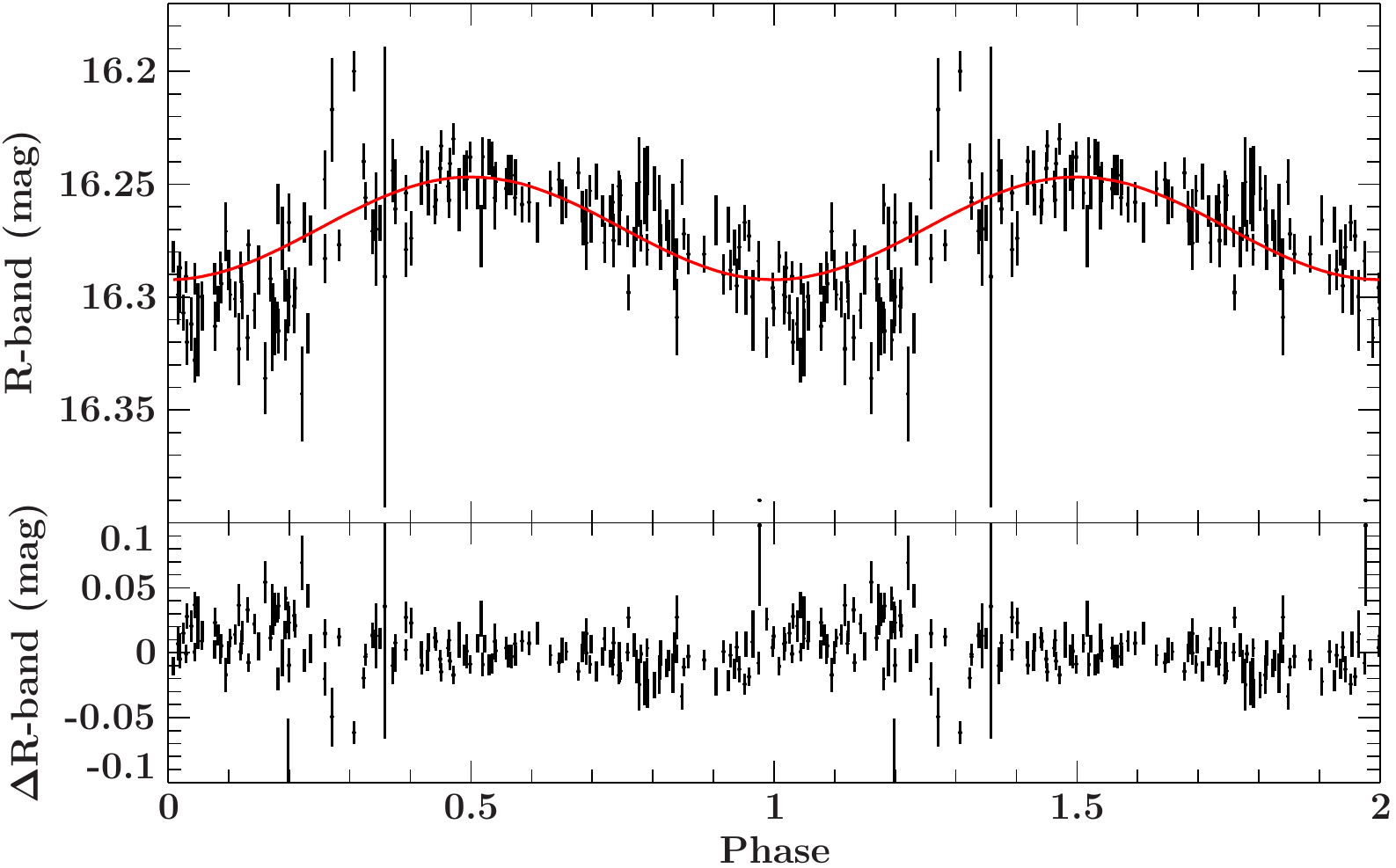}
\caption{{\it Upper}: folded PTF lightcurve of J1250 at the best-fit period, $P=2.92217$ d, together with a sine fit. For ease of comparison of the photometric and spectroscopic data, the sine wave shown here uses the best-fit photometric period and amplitude, but the best-fit spectroscopic phase. {\it Lower:} residuals from the fit.}
\label{fig:lightcurve}
\end{center}
\end{figure}  

We searched for shorter term (minutes) photometric variability in J1250 via observations with the Palomar 1.5m telescope on multiple nights in 2017. No significant rapid variations were observed.

There is only one previous mention of J1250 in the literature, where G13 first notes it as an SDSS dC star, with a distance estimate of 300 pc.  He also remarks on two unusual properties: the star displays strong Balmer emission in the SDSS spectrum (rare, but not unprecedented in the known sample of dC stars), as well as very weak soft X-ray emission detected by XMM (unique in the G13 sample, but consistent in luminosity with observations of chromospheric/coronal activity in other late type stars. 

J1250 appears in the Catalina Real Time Transient Survey (CRTS) \citep{2009ApJ...696..870D}, where no significant photometric periodicity is noted. Given the low amplitude of the periodicity that we observe in PTF, this non-detection in CRTS at this magnitude is not surprising.

Limited other photometric measurements are available in the literature. J1250 has JHK photometry available in 2MASS \citep{2006AJ....131.1163S}.  The object is not detected by GALEX \citep{2014AdSpR..53..900B}.

\subsection{Spectroscopy}
We obtained six spectra of J1250 on 2017 April 20, 21, 22, using the Echellette Spectrograph and Imager (ESI) \citep{2002PASP..114..851S} at the Keck II telescope. This instrument covers the 3,900-10,000 \AA\,  range, and the 0.75\arcsec~slit employed yields resolution $R=7000$.  A typical exposure time was 600~s.  We also obtained lower resolution spectra on three additional nights in 2017 using the Palomar 5m reflector and the Double Beam Spectrograph \citep{1982PASP...94..586O}. The Keck and Palomar spectra are similar to the SDSS discovery spectrum, displaying deep $C_2$ Swan bands, strong (resolved) NaD doublet absorption, the red CN bands, and prominent Balmer and Ca H\&K emission. At the substantially higher signal to noise and spectral resolution of the Keck data, numerous other cool star absorption features are also clearly detected, including multiple  Ca\,I and Ca\,II lines. The Keck spectra resolve the Balmer emission; H$\alpha$ has 0.6 \AA\, FWHM. Comparison of spectra on different nights makes it immediately apparent that there are substantial velocity shifts. The Balmer emission velocities in each spectrum agree to within the measurement errors with the velocities of multiple Ca\,{\sc i}, Ca\,{\sc ii}, and NaD absorption lines, so the emission and absorption appear to come from similar locations. In the subsequent analysis we use the multiple strong, sharp Balmer emission lines in the Keck spectra for velocity measures, due to their very high signal to noise ratio. The Palomar spectra yield consistent velocities, but are not included in the analysis, as their substantially lower resolution fails to resolve the emission lines.

\begin{figure}
\begin{center}
\includegraphics[width=0.49\textwidth]{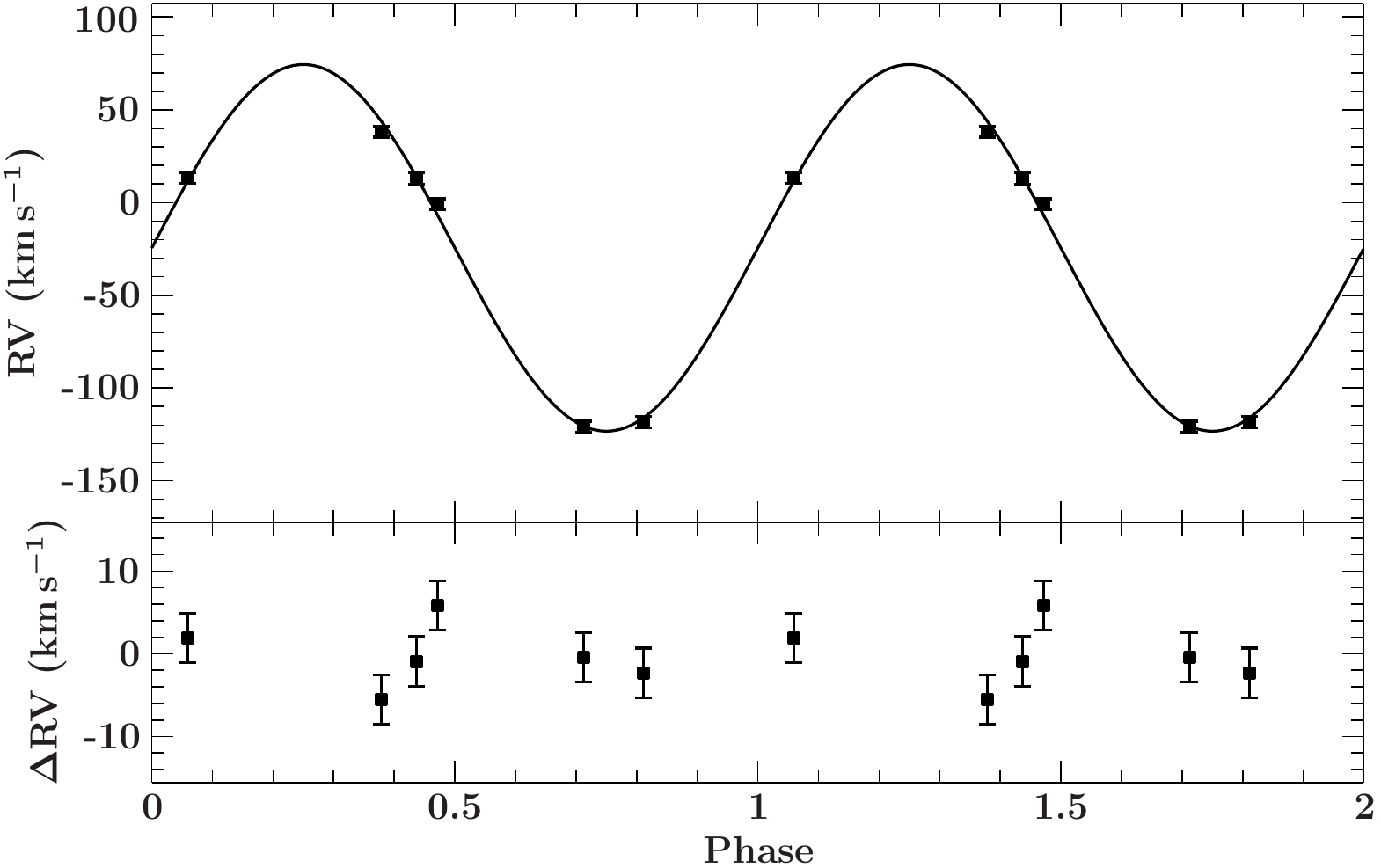}
\end{center}
\caption{Radial velocity of J1250 vs. orbital phase, from Keck/ESI spectra. The RV data are folded with the photometric orbital period and plotted twice for clarity, with residuals shown below.}
\label{fig:rv_curve}
\end{figure} 

\begin{table}[t]
 \centering
 \caption{ESI spectra of J1250}
  \begin{tabular}{cllll}
  \hline\hline
  HJD 2457860 +   &   Phase  &  Velocity &   H$_\alpha$ EW & FWHM \\
   (d)  &         &    km~s$^{-1}$   &   \AA  &  \AA  \\
  \hline
4.763347 & 0.37875  & $38.2\pm3.1$   &  $5.2\pm0.4$  & $0.60\pm0.03$  \\
4.932885 & 0.43668  & $12.9\pm3.2$   &  $5.1\pm0.4$  & $0.61\pm0.03$  \\
5.034631 & 0.47145  & $-1.0\pm3.0$   &  $5.4\pm0.4$  & $0.68\pm0.03$  \\
5.739630 & 0.71234  & $-121.0\pm2.9$ &  $5.6\pm0.4$  & $0.64\pm0.03$  \\
6.029278 & 0.81131  & $-118.4\pm3.1$ &  $5.5\pm0.4$  & $0.63\pm0.03$  \\
6.753926 & 0.05891  & $13.2\pm3.1$   &  $5.8\pm0.4$  & $0.59\pm0.03$  \\
   \hline
\end{tabular}
\label{tab:ESI}
\end{table}

The PTF photometric period determination has excellent precision due to the long time baseline. We have therefore folded the radial velocities determined from the Keck spectra with the best fit PTF 2.92~d photometric period, and fit these folded data with a sine wave, allowing the amplitude and phase to remain as free parameters. The folded data and resulting best fit are shown in Figure\,\ref{fig:rv_curve}; the sine function is clearly a good fit, with velocity half-amplitude $K=98.8\pm10.7$~km~s$^{-1}$ and systemic velocity $\gamma=-24.7\pm4.8$~km~s$^{-1}$. Phase zero at HJD$=2457863.6548\pm0.001$\,days corresponds to when the dC is closest to us.

\section{Discussion}

 Although there are composite spectrum dC stars known, we believe this to be only the second spectroscopic binary, finally complementing the prototype dC, G77-61. However, the parameters of this system are clearly very different. The observed radial velocity variations and 2.92~d period imply a mass function $f(M)=0.29$\,\msol. The colors and spectrum of J1250 (with the exception of the $C_2$ bands) suggest that it is of the (most common) $K$-type dC. For example, the observed prominent CN bands and Balmer emission are not found in the rarer, warmer $G$-types (G13), and the latter also fall in a restricted SDSS color wedge that differs from the observed J1250 colors. The mass of the dC is uncertain, but adopting the dK assumption, is likely $\sim 0.3$~\msol, implying a minimum companion mass $M_{comp}\sim$ 0.6~\msol. The lack of detectability of the companion in the visible spectrum, plus the requirement that it be highly evolved to have donated the observed $C_2$ on the dC, indicates a compact object, and the inferred $M_{comp}\sim$ 0.6~\msol\, is consistent with a typical white dwarf mass. Given the surprisingly short orbital period of 2.92~d, substantial orbital evolution must have occurred.

The existence of radial velocity variations together with the observed photometric and radial velocity phases rules out a variety of models for the periodic brightness modulation, such as pulsation or ellipsoidal variations, and suggests that the unseen white dwarf is still sufficiently luminous to heat the face of the dC. The observed Balmer emission might be an additional indicator of the heating, although the data in Table 1 indicate there is no obvious variability of the emission line strength with binary phase. Alternatively, the emission may well be related to stellar activity, as also evidenced by the faint X-ray emission (G13).

\section{Conclusion}
More than 30 years after \citet{1986ApJ...300..314D} reported G77-61, the prototype dC star, as a 245~d spectroscopic binary, J1250 finally presents a second example, as well as the first dC with synchronous photometric variations. The two objects are clearly quite different.  The simulations of \citet{1995ApJ...449..236D} predict a bimodal orbital period distribution for dC's, centered on periods of a few years and a few decades.  We have considered the possibility that J1250 is a triple system, with an as yet undiscovered donor star in a wide orbit, as has been suggested for SDSS J1837+40 by G13. However, we know that the observed close companion of J1250 is itself a subluminous, evolved star of the type hypothesized in all dC's.  It thus seems unnecessary to invoke an additional, undiscovered subluminous object.

For the sake of completeness, we note that the relationship of dC stars and the better-studied carbon-enhanced metal poor (CEMP) stars (e.g., \citet{2007ApJ...655..492A}) is unclear. Although the latter are mostly giants and subgiants, and they are certainly selected via techniques quite different from dCs, there are a handful with $C_2$ bands where model calculations indicate high gravities indicative of dwarfs. Thus these objects are inferred to be dwarfs via modeling, rather than direct observation of parallaxes or proper motions as is the case for dC stars. At least one of these, HE 0024-2523 \citep{2002AJ....124..481C}, often referred to as a CH or even Pb star, is known to be a spectroscopic binary with $P=3.14$\,d and $K = 50$\,\kms\, \citep{2003AJ....125..875L}. It may well be the case that the dC and high-gravity CEMP stars are the precursors of the CEMP giants.

Because the orbital periods of G77-61 and J1250 are so different, it is quite unclear how many shorter period dC systems such as J1250 exist and await discovery via photometric variations such as observed here. Current and future large scale synoptic photometric surveys, coupled with multi-color candidate selection such as that employed by SDSS, may well prove a fruitful discovery mechanism.

\section*{Acknowledgments}
We are grateful to P. Green for numerous conversations about dC stars, and thank E. Ofek for comments on the manuscript. Observations obtained with the Samuel Oschin Telescope at the Palomar Observatory as part of the Palomar Transient Factory and Intermediate Palomar Transient Factory project, a scientific collaboration between the California Institute of Technology, Columbia University, Las Cumbres Observatory, the Lawrence Berkeley National Laboratory, the National Energy Research Scientific Computing Center, the University of Oxford, the Weizmann Institute of Science, Los Alamos National Laboratory, the University of Wisconsin, Milwaukee, the Oskar Klein Center,  the TANGO Program of the University System of Taiwan, and the Kavli Institute for the Physics and Mathematics of the Universe. Funding for the SDSS and SDSS-II has been provided by the Alfred P. Sloan Foundation, the Participating Institutions, the National Science Foundation, the U.S. Department of Energy, the National Aeronautics and Space Administration, the Japanese Monbukagakusho, the Max Planck Society, and the Higher Education Funding Council for England. The SDSS website is http://www.sdss.org/. Some of the data presented herein were obtained at the W.M. Keck Observatory, which is operated as a scientific partnership among the California Institute of Technology, the University of California and the National Aeronautics and Space Administration. The Observatory was made possible by the generous financial support of the W.M. Keck Foundation. The authors wish to recognize and acknowledge the very significant cultural role and reverence that the summit of Maunakea has always had within the indigenous Hawaiian community. We are most fortunate to have the opportunity to conduct observations from this mountain.

\facility{Keck:II (ESI), Sloan, Hale, PO:1.2m, PO:1.5m}


\end{document}